# Intellectual Up-streams

# of Percentage Scale (*ps*) and Percentage Coefficient ($b_p$)

# -- Effect Size Analysis (Theory Paper 2)

**Authors:** Xinshu Zhao (xszhao@um.edu.mo)[1], Qinru Ruby Ju (qinrujv@126.com)[2], Piper Liping Liu (llpsxx@hotmail.com)[3], Dianshi Moses Li (yc37228@um.edu.mo)[2], Luxi Zhang (yc27303@um.edu.mo)[1], Jizhou Francis Ye (yjz199745@gmail.com)[4], Song Harris Ao (harrisao@um.edu.mo)[5], Ming Milano Li (yc17316@um.edu.mo)[1]

**Affiliations:**

[1]Faculty of Social Sciences, University of Macau; Macau SAR, 999078, China;

[2]Centre for Empirical Legal Studies; Faculty of Law; Macau SAR, 999078, China

[3]School of Media and Communication, Shenzhen University, China

[4]Department of Communication, University of Oklahoma, Norman, Oklahoma, USA

[5]School of Journalism and Communication, Sun Yat-sen University, China

**Abstract**

Percentage thinking, i.e., assessing quantities as parts per hundred, has traveled far and wide from Roman tax ledgers to modern algorithms. Building on early decimalization by Simon Stevin in La Thiende (1585) and the 19th-century metrication movement that institutionalized base-10 measurement worldwide (Cajori, 1925), this article traces the intellectual trails through which base-10 normalization, especially 0~1 percentage scale, trudged to become a unifying language for both human and machine understanding. We retrack the baby steps researchers of UW-Madison and UNC-Chapel Hill took from the 1980s to "percentize" scales to make regression coefficients interpretable. We discuss the commonalities between the Wisconsin-Carolina experiments and the established indices, especially the ±1 Pearson (1895) correlation (*r*) and 0~1 coefficient of determination, aka *r* squared (Wright, 1920). We pay tribute to the influential *percent of maximum possible* (POMP) coefficient by Cohen et al. (1999).

The history of the 0~100 or 0~1 scales goes back far and wide. Roman fiscal records, early American grading experiments at Yale and Harvard, and contemporary analysis of percent scales (0~100) and percentage scales (0~1, or -1~1) show the tendency to rediscover the scales and the indices based on the scales (Durm, 1993; Schneider & and Hutt, 2014).

Data mining and machine learning since the last century adopted the same logic: min-max normalization, which maps any feature to [0, 1] (i.e., 0-100%), is predicted to equalize the scale ranges, thereby equalizing the weights of the predicting variables, maximizing the predictive accuracy. Because 0~1 percentage scale assigns the entire scale to be the unit, equalizing the scales also equalizes the units of all percentized scales. Having equitable units of scales is the necessary and sufficient condition for comparability of two indices, according to the percentage theory of measurement indices (Cohen et al., 1999; Zhao et al., 2024; Zhao & Zhang, 2014). Thus, the massive successes of machine learning and artificial intelligence of the last half century serve as gigantic technical and social experimentations confirming the comparability of the percentage-based indices, foremost among them the percentage coefficient ($b_p$).

**Introduction**

From ancient commerce to modern algorithms, the idea of expressing quantities “per hundred” has long served as a common language for understanding and communicating numbers. Traces of percentage thinking date back to Roman taxation (Gutiérrez & Martínez-Esteller, 2022; Scheidel & Friesen, 2009). The adoption of decimal fractions during the Renaissance unlocked much of its potential. Mathematicians like Simon Stevin popularized base-10 calculations in the 1500s, comparing values as parts of 100 natural (Clark, 2009; Struik, 1959). But it was the metrication movement that started in the 19th century that encouraged scientists and educators to measure diverse phenomena on unified scales, fostering a decimal and numerical mindset (Sarkar & Salazar-Palma, 2016; Wolfle, 1965). By the late 1800s, even academic grading had begun to use a 0-100 scale, as universities such as Yale and Harvard experimented with percentage-based marks for performance evaluation (Durm, 1993; Schneider & and Hutt, 2014). These developments situated the percentage scale within the broader intellectual history as a tool for clarity, standardization, and comparability across measurements.

In parallel, the social sciences started harnessing percentage scales to quantify intangible constructs. Psychologists and sociologists found that recasting survey scores or index values onto a 0–100 range made results more interpretable for researchers and laypersons (Cohen et al., 1999). For example, a response mean of 7.5 on a 0–10 attitude scale could be intuitively reported as 75%, directly conveying its position relative to the scale’s maximum. By the late twentieth century, scholars at the University of Wisconsin and the University of North Carolina at Chapel Hill began to "percentize" psychological test scores and socioeconomic indices as a part of regression or ANOVA analysis. These experiments underscored how converting disparate measures onto a common percent scale could reveal patterns that raw scores obscured. This era established a dual root for percent metrics in social science—as both a communication tool and an analytic strategy for achieving conceptual anchoring (defining 0 as absence of a trait and 100 as full presence) across different measures.

A separate but convergent stream was unfolding in computing and data science. As algorithms grew sensitive to the scale of input data, normalization techniques emerged to put variables on comparable footing. A seemingly simple and now-regular approach is min–max rescaling, which maps any variable's range to [0,1] (often reported as 0–100%). This practice, common in machine learning, treats each feature, aka variable, as a percentage of its "true" range — echoing social scientists' percentization that is based on "conceptual" range. By the late 20th century, it was well understood that machine learning models perform best when inputs and outputs are on normalized scales. Notably, model performance metrics themselves are frequently expressed in percentage terms: classification accuracy, for instance, is simply "percent correct (Congalton, 1991; Fawcett, 2006)." These are independent discoveries and rediscoveries of percentage scaling to achieve interpretability, comparability, and numerical stability. While social scientists percentized, aka normalized, survey and experimental data to interpret human behavior, computer scientists normalized, aka percentized, various types of data, especially natural big data, to tame algorithms. Without extensive communication between them initially, the two groups discovered and cultivated one same treasure, which social scientists named “percentage scale.”

Most recently, the conceptual foundations of percentage metrics have been unified under a new theoretical lens. Zhao’s two-function theory posits that statistical indicators must fulfill two primary purposes: to aid comprehension and to enable comparison (Zhao et al., 2024). A percentage scale naturally excels at both. On one hand, it offers clear interpretability – a score of 20% vs. 80% immediately conveys a low vs. high position on a construction. On the other hand, because any percentage is measured on an identical 0–100 framework, it allows equitable comparisons across different variables, populations, or studies. In line with this theory, recent work has defined the percentage coefficient, denoted (bp), as a novel effect size for regression models (Jiang et al., 2021). This coefficient is computed by first converting the independent and dependent variables to conceptual 0–1 (0–100%) scales and then estimating the slope. The result is an effect size with a tangible meaning: it represents the expected change in the outcome in percentage points for a full-range (100-point) increase in the predictor. Such an indicator directly embodies the twin functions of comprehension and comparison – it is immediately interpretable and directly comparable across contexts. By integrating ideas from social science measurement and computer science normalization, Zhao and colleagues have given formal voice to the intuitive practice of percent-based reasoning in research.

This paper follows these intellectual upstreams to their confluence. We trace how the idea of percentage scale and percentage coefficient, aka percentage thinking, evolved from disparate origins into a coherent framework of concepts, principles, and techniques. We retrack the historical path from decimalization to percentization and explore the rationale for anchoring scales first at [0, 100] and later at [0, 1] or [-1, 0, 1]. We review the theory that percent-based metrics enable interpretability and enhance comparability.

In doing so, we highlight the importance of conceptual anchoring, i.e., 1) defining a ground zero that is truly an *absolute zero* or *true zero* as required by the classic theory of measurement scales (Stevens, 1946, 1951); 2) defining a pair of *conceptual min-max* that may differ from *observed min-max* and may further differ from the *possible min-max*. We emphasize that interpretability requires uniformly meaningful scale units; comparability requires equitable units; uniform meaningfulness and equitable units both require proper ground zero and conceptual min and max.

This journey of discovery and refinement underscores a larger theme: the power of rigorous simplicity. Percentization, the conversion of measurements into percentage metrics, emerges not as a mere convenience but as a disciplined strategy to achieve clarity and normality, i.e., clear meanings and meaningful comparisons, in research. In the chapters that follow, we chronicle this evolution and discuss how embracing percentage scales and the $b_p$ coefficients can lead to more interpretable, theory-anchored, and generically comparable findings. The convergence of historical insight with contemporary practice offers a richer understanding of why 0~1 and -1~1 based indices are on their way to becoming platform on which diverse scientific stories can be reported, revised, and reconstructed.

By revisiting key steps like the Wisconsin–Carolina experiments, the POMP framework, the min-max normalization for feature scaling and machine learning, and the two-function theory, we reveal a narrative of convergent evolution by three groups, a small team of psychologists, another small team of social scientists, and many computer scientists and data miners. Through their work, what began as a convenient counting gimmick by ancient hunters, farmers, and merchants has evolved into a conceptual viewfinder data miners use to make artificial intelligence more intelligent, and social scientists use to make themselves more intelligent, connecting domains as far apart as early social experiments and cutting-edge machine learning. Given time and luck, it – we mean percentage thinking – may evolve into a language by which scientists and other people communicate their understandings of world, making humankind more intelligent and more capable of managing artificial intelligence.

# I. Intellectual Up-streams of Percentage Scale and Percentage Coefficient ($b_p$)

Percentage coefficient ($b_p$) is a regression coefficient when the dependent variable (DV) and the independent variable (IV) are each on a 0~1 percentage scale. While appearing straightforward, the scale can be traced to long and varied up-streams.

## I.1. Decimal numerals and metric systems

Possibly because human ancestors used ten fingers to help count, number 10 and its powers, 1, 10, 100, 1,000, etc., and 0, seem to play special roles in our brain struggling to make sense of the world. These numbers have special characteristics. Multiplying any real number with 0 returns 0. Multiplying any real number with 1 returns the number. Multiplying 10 or 100 simply moves the decimal point to the right, while dividing 10 or 100 just moves the point to the left.

The best-known ancient civilizations, including Brahmi, Chinese, Egyptian, Greek, Hebrew, Roman, and all used the decimal systems, which are based on the number ten and its powers (Lockhart, 2019) (Wikipedia, 2025b). The Hindu-Arabic numerals that dominate today's human counting worldwide is also a decimal system. Originating in India in the $6^{th}$ or $7^{th}$ century, it was introduced to Europe about the $12^{th}$ century through the work of al-Khwarizmi and al-Kindi, among others (Britannica, 2024a).

The influence of finger counting and 10-based numbering on human thinking is also reflected in the etymology and semantics of three common words. The word "digital," which means numerical, electronic, computerized, virtual, or online in modern English, originated from the Latin root *digitālis*, which meant a finger's breath or relating to the finger (OED, 2025b). The word "decimal," which can mean a fraction of one in modern English, originated from the Latin root *decimus* and Sanskrit *daśamá,* meaning tenth (OED, 2025a).

The third word, "metric," which came from French métrique and mètre and was traced to Greek μέτρον, meant just "relating to measure or measurement" before the 18th Century (OED, 2024, 2025c). In 1791, however, the French National Assembly, amidst the French Revolution (1789~1799), defined "mètre" (British English metre and US spelling meter, denoted "m") to be one ten-millionth of the distance from the equator to the North Pole (International Bureau of Weights and Measures, 2019; Wikipedia, 2025c). It was the base-level unit of a decimal system of length, with kilometer (1,000m), centimeter (1/100 m), and millimeter (1/1,000 m) being the other often-used levels of units. The metric system of mass, i.e., weight, had gram (g), kilogram (1kg=1,000g), metric ton (1t=1,000kg), and milligram (1mg=1/1,000 g) as the levels of units. As length and weight are among the most used measurements in everyday life, for many people throughout the world the word "metric," as the key word in "metric system," as taken on the meaning of "relating to international and decimal system," where "decimal" means "relating to the number ten, the tenth parts, or powers of ten." (OED, 2025a, 2025c)

It was John Wilkins who proposed in 1668 to apply decimal scales to measure length and weight (Naughtin, 2012; Rooney, 2012). During the Age of Enlightenment (1685-1815), the idea evolved into a metric system of kilogram and metre (Britannica, 2024b; Giunta, 2023; Maestro, 1980; Wikipedia, 2025a). The metric system was officially adopted in France in 1799, during the French Revolution. The process of adopting and applying the metric system, known as metrication or metrification, was complete in France and much of Europe by the 1850s (Wikipedia, 2025a, 2025e). After the Treaty of the Metre in 1875 and the adoption of the International System of Units in 1960, the metric system is now used officially in almost all countries in the world (Britannica, 2024b). The measurement systems under the umbrella of "metric" measure not just length and weight, but also volume, time, temperature, electricity, substance, and light (Wikipedia, 2025d). Given time, would a decimal system of effects become imaginable?

### I.2. Decimal Scale, Percentage Scale, and Percentage Coefficient

The metric system's successes are often attributed to the decimal system of counting that enables and eases comprehension and comparison more effectively than any other system (Maestro, 1980) (Giunta, 2023). Regression researchers also cited the two features of *percentage scale* and *percentage coefficient*, which are two decimal scales measuring variables and effects (Zhao et al., 2024; Zhao & Zhang, 2014).

Ranging conceptually 0~1 or -1~0~1, percentage scales constitute a unique class of decimal system (Zhao, 2008, 2010; Zhao & Zhang, 2014). Here, the word "decimal" carries two meanings. One is about numeral ten. The range 0~1 or -1~0 is one-tenth (1/10) of ten. The other is about decimal points. As percentage scales range between -1, 0, and 1, the values in between must be represented by decimal points.

They are named *percentage scales* because moving the decimal point two positions to the right would turn a 0~1 scale to a 0~100 scale. A figure 0.x is routinely interpreted as x%, e.g., 0.87=87%. This gives us four fundamental features of percentage scales:

Feature 1: $$0.x = x\% \quad (1)$$
Feature 2: $$0 \cdot x = 0 \quad (2)$$
Feature 3: $$1 \cdot x = x \quad (3)$$
Feature 4: $$-1 \cdot x = -x \quad (4)$$

The features empower the scales. Pearson's correlation coefficient (*r*) features a bidirectional percentage scale ranging from -1~0~1, making it among the first applications of the percentage scale in correlation analysis (Bravais, 1844; Galton, 1877, 1885, 1886, 1889; Pearson, 1895). The success of Pearson's *r* has been attributed in part to the percentage scale that enables and encourages interpretation and comparison (Rodgers & Nicewander, 1988; Wright, 1921). In parallel, the success of the *coefficient of determination*, more commonly known as $R^2$ or $r^2$, has been attributed to its 0~1 percentage scale (Glantz et al., 2016; Steel & Torrie, 1960; Wright, 1920, 1921).

### I.3. Wisconsin-Carolina Experiments with 0~100 Scale and First $b_p$ Coefficients

The experiments began in the 1980s to make regression coefficients interpretable and comparable by placing variables on 10-based scales. In a series of semi-natural experiments and structured observations, communication and consumer researchers at the University of Wisconsin-Madison placed dependent variables (DV), including memory and liking of advertised brands, on 0~100 scales using Eq. 5) for regression analysis (Hitchon et al., 1988; Thorson et al., 1987; Thorson & Zhao, 1989, 1997/2014; Thorson et al., 1988; Zhao, 1989).

$$s_n = \frac{s_o - min}{max - min} \cdot 100 \qquad (5)$$

Where --
$s_n$: new score after transformation.
$s_o$: original score on original scale.
$min$: minimum on the original scale.
$max$: maximum on the original scale.

In a questionnaire survey, Zhao and Xie (1992) placed their main DV, attitude toward ideology, on a 0~100 scale for ANOVA and regression analysis. Zhao, Zhu, Li and Bleske (1994) did the same in another cross-sectional survey, placing their four DVs, including one knowledge measure and three attitude measures, on 0~100 scales. As the practice was uncommon, the authors explained repeatedly that the scale transformation was "to ease interpretation in regression analysis" (pp. 100-101). When the studies began, Zhao and Xie were doctoral students at University of Wisconsin–Madison. When the studies were published, Zhao had taken a teaching position at the University of North Carolina at Chapel Hill, where Bleske was a doctoral student. Zhao and UNC students also applied the 0~100 scale to analyze data from a naturalist quasi-experiment with content-coded IV and survey-measured DV (Jeong et al., 2011; Jin & Zhao, 1999; Jin et al., 2006; Kim & Zhao, 1993; Zhao et al., 1993).

In the 1994 study, the independent variable *age* had years as the unit (Zhao et al., 1994). Thus, age was in effect on a 0~100 scale with 100 being the conceptual maximum, which later studies converted to 0~1 scales (Peng et al., 2020; Zhao et al., 2023; Zhao & Zhang, 2014). When DV and IV are on the same ratio scale, regression produces a coefficient now known as *percentage coefficient* ($b_p$) (Zhao et al., 2024; Zhao & Zhang, 2014). Thus, the 1994 study produced four $b_p$ coefficients, with *knowledge* and three *attitude* variables as the DVs and *age* as the IV (Table 3 of Zhao et al., 1994). They are the first $b_p$ coefficients known to have been published.

The four coefficients are also the first published POMP coefficients that Cohen et al. (1999) recommended, because POMP and $b_p$ coefficients are numerical equivalents when the IV is on a numerical scale.

The 1994 study also conducted a mediation analysis between the four variables. As all four are on the same scale (0~100), the unstandardized path coefficients were equivalent to POMP coefficients (Cohen et al., 1999) and to the percentage coefficients, $b_p$ (Zhao & Zhang, 2014). However, without the guidance of the POMP theory (Cohen et al., 1999) or the functionalist percentage theory (Zhao et al., 2024; Zhao & Zhang, 2014), the mediation path diagram still used standardized coefficients (β) even though the $b_p$ coefficients were also calculated for the mediation and path analysis (Zhao et al., 1994).

In the 1994 study, the normalized scales also enabled the researchers to compare effects across the dependent variables and test their post-hoc theory of media messages being relayed through cognitive and attitudinal nodes, like signals through watchtowers of the Great Wall and signal strengths reduced through the nodes/watchtowers. The same model would be referred to as serial mediation in the 2020s. Thus, the 1994 study is also the first known application of normalized scales in *relative impact* analysis, which compares the impacts of the same IV on different DVs.

Zhao (1997) featured another application of the 0~100 scale. The study measured *liking* using a 7-point Likert in two years (1992 & 1994) and a 9-point Likert in one year (1993). By converting the *liking* of each year to a 0~100 scale, the study was able to pool data from different years to make liking a unified variable. By also placing two other DVs, recall and recognition, on 0~100 scales, the study was able to compare the *relative impacts* on the three DVs. Adopting this tool, Youn et al. (2001) pooled data from two locals, Chapel Hill – Carrboro, NC, and Twin Cities,

MN, and three different years, 1995, 1996, and 1997, to compare relative impacts on memory and liking.

Zhao and Bleske (1998) study may be the first application of 0~100 percent scales in a controlled experiment. After converting their DV onto a 0~100 percent scale, the researchers assessed the percentage difference in DV between the treatment and control groups. This *percent difference* is based on a 0~100 percent scale anchored by a conceptual maximum and a conceptual minimum; therefore may be seen as a concept-based alternative to the variation-based *Cohen's d* (1988).

**I.4. Machine Learning, Feature Scaling, and Min-Max Normalization**

At the turn of the century, feature scaling and scale normalization became a routine operation in data mining and machine learning (Han & Kamber, 2001; Han et al., 2012; Jain et al., 2005; Juszczak et al., 2002). Foremost among them, min-max normalization uses Eq. (6) to equalize the range of IVs predicting a DV (Shalabi et al., 2006). IVs of larger ranges exert heavier weights, and DVs of larger ranges receive lighter weights. As ranges of raw scales may differ from each other by millions or more times, untreated scales imply drastically different weights between variables, aka features. As machine learning and artificial intelligence (AI) are set to import all information from all features available, unequal weights between variables would demolish the foundation and beat the purpose of aiding making.

Min-max normalization, as shown in Eq. (6), is to normalize the raw scales to equalize the weights of features (Han et al., 2012, p.114; Shalabi et al., 2006, p.735). A programmer can set the transformed scale to a chosen range by setting the values of $min_n$ and $max_n$. The often-used ranges include 0~1, -1~1, and 0~100 scales.

$$v' = [\frac{v - min_o}{max_o - min_o}(max_n - min_n)] + min_n \qquad (6)$$

$$(min_n \leq v' \leq max_n)$$

$v'$: new score after scale transformation
$v$: original score before scale transformation
$max_o$: maximum on original scale
$min_o$: minimum on original scale
$max_n$: maximum on new scale
$min_n$: minimum on new scale

Computer and data scientists consider scale normalization a part of data *preprocessing* (Shalabi et al., 2006). Introductory texts would detail the procedure and focus on application. To computer scientists, data scientists, and AI developers, it was self-evident that equalizing the variables' range equalizes the features' weights. To us and other social scientists, the phenomenon success of machine-learning and artificial intelligence (AI), which are all based on normalized scales, add countless real-data confirmations. That is, scale normalization equalizes the scales' units, thereby enabling comparison between the variables' efficiencies measured by their coefficients.

### I.5. Cohen et al's POMP Theory and 0~100 Percent Scale

Cohen et al. (1999) advanced a theory stating that scale units must be meaningful for the statistical indicator to be meaningful. This is an important step in advancing our understanding of effect sizes. The authors proposed converting close-ended ordinal scales to 0~100 scales for calculating the regression coefficient, which they named *the percent of maximum possible score* (POMP). We take the word "percent" to name the 0~100 scale "percent scale," as supposed to 0~1 "percentage scales."

### I.6. Converging of Thoughts and Practices

While the three groups of researchers operate independently, their work complemented each other. Cohen et al. (1999) and the afore cited Zhao and colleagues (1987-1998) both used Eq. 5 for the conversion to the 0~100 scales. The computer scientists used the same equation 5 when they set $min_n$=0 and $max_n$=100 in Eq. 6). Independently, Cohen et al. (1999) provided theoretical justification, computer scientists provided big-data verification (Han & Kamber, 2001; Jain et al., 2005; Juszczak et al., 2002), and the Wisconsin-Carolina researchers provided social science applications (Thorson et al., 1987; Zhao, 1989; Zhao & Bleske, 1998; Zhao et al., 1994).

### I.7. Percentage Coefficient ($b_p$) on 0-1 Scale and Two-Function Theory

When reanalyzing the data of Zhao (1997), Zhao et al (2010) replaced the 0~100 percent scale they had used since 1987 with a 0~1, which is now known as *percentage scale* (PS). To do so, they replaced Equation 5 with Equation 7. Placing a variable on a 0-to-1 or 0-or-1 scale is now known as *percentization*.

$$s_p = \frac{s_w - c_n}{c_x - c_n} \qquad (7)$$

Where --

$s_p$: *Percentage* score after transformation.

$s_w$: Raw score before transformation.

$c_n$: Conceptual minimum on the raw scale before transformation.

$c_x$: Conceptual maximum on the raw scale before transformation.

Percentization again allowed the researchers to equalize the scales for their DV, Liking, which had been measured using different scales in different years. But the 2010 study did not percentize the IV. Therefore, the study did not produce a percentage coefficient ($b_p$) based on 0-1 percentage scale. That would have to wait for four more years.

Zhao and Zhang (2014) define two functions for statistical indices: 1) assisting interpretation of target phenomenon, and 2) assisting comparison between phenomena. To fulfill the first mission, the units of the measurement scale must be meaningful (First Requirement). To fulfill the second mission, the units of the scales under comparison must be equivalent to each other (Second Requirement). While a 0~100 percent scale meets both requirements when independent variables (IV) are all numerical, it does not satisfy either when one or more IV is dichotomous and/or nominal coded as 0-or-1 binary variables.

Accordingly, Zhao and Zhang (2014) suggested converting variables -- numerical, nominal, or dichotomous -- to 0-to-1 or 0-or-1 *percentage scales* ($p_s$).

Later, to compare numerical IVs with dichotomous IVs on 0-or-1 dummy scales, the 0~100 scales were switched to 0~1 scales using a formula equivalent to Eq. 7 (Zhao et al., 2010; Zhao & Zhang, 2014). Zhao and Zhang (2014) recommended transforming all variables – DV, IV, and control variables, numerical and dummy variables – to 0-1 scales. They coined the term *percentage scale* to represent 0-1 scales, including 0-or-1 dummy scales and 0-to-1 numerical scales, and the term *percentage coefficient* ($b_p$) to represent the regression coefficient on such scales.

They coined the term *percentage coefficient* and the symbol $b_p$ to represent the regression coefficient when DV and IV are both on percentage scales. They reanalyzed the data of Zhao et al. (1994) as an example.

Peng et al.'s (2020) empirical study applied $b_p$ as the main effect size indicator. Jiang et al.'s (2021) mathematical proof, published in a journal of mathematical statistics, applied $b_p$ to showcase its application in two empirical examples.

**I.8. Comparisons Across Scale Types**

Early developers of percentage scales faced two challenges: 1) placing numerical, binary, and nominal measures on comparable scales, and 2) placing close-ended and open-ended measures on comparable scales. Without a solution to either challenge, Zhao et al. (1994) and Cohen et al. (1999) were unable to conduct most of the comparisons between the effects of two or more independent variables (IVs), which trail blazers of modern statistics referred to as *relative importance* analysis (Blalock Jr, 1961; Davenport, 1917; Wright, 1918, 1920). These comparisons include the following:

1) Comparison between numerical IV and binary IV.

2) Comparison between numerical IV and nominal IV.

3) Comparison between binary IV and nominal IV.

4) Comparison between two nominal IVs, each having a different number of categories.

5) Comparison between closed-end numerical IV and open-end numerical IV.

6) Between two open-end numerical IVs on different scales with different units.

The studies of the 1980s and 1990s also did not conduct any comparison between effects on two or more dependent variables (DVs), now known as *relative impact* analysis (Cohen et al., 1999; Thorson et al., 1987; Zhao, 1989; Zhao et al., 1994). At the time, neither the concepts nor the techniques were ready for a comprehensive analysis of relative impact. The comparability based on the 0~100 percent scales or POMP scales was restricted to between close-end numerical scales, not extendable to any other scales.

To meet the first challenge, Zhao and Zhang (2014) switched from 0~100 scales to 0~1 scales for numerical variables, so that all variables, numerical, nominal, and binary, may share the same range, 0-to-1 or 0-or-1. To meet the second challenge, the authors proposed what we call *conceptual anchoring* and *neighborhood rounding*, as discussed below.

### *I.8.1 Conceptual Anchoring and Neighborhood Rounding*

To meet the second challenge, Zhao and Zhang (2014) developed two related procedures, *conceptual anchoring* and *neighborhood rounding.* Conceptual anchoring means to choose scale maximum and scale minimum based on conceptual legitimacy and appropriateness, but not necessarily on the appearance of the data in hand or mathematical derivations and theorems. The authors used the term "anchor" to explain their procedure, and they reanalyzed data published 10 years earlier to illustrate the procedure (Zhao et al., 1994). Using variable *age* as an example, Zhao and Zhang (2014) selected 0 as the *conceptual minimum* and 100 as the *conceptual maximum,* even though the *observed minimum* was 18 and the *observed maximum* was 83. After percentization using Eq. 8, which was derived from Eq. 7, the new age variables had an observed range of 0.18~0.83 and a conceptual range of 0~1.

$$new\ age = \frac{raw\ age - 0}{100 - 0} \quad (8)$$

### *I.8.2 Neighborhood Rounding*

Conceptual anchoring defines a measurement scale by posts or pillars that researchers select based on conceptual appropriateness and theoretical soundness, rather than positions or points that emerged from the data under examination. *Neighborhood rounding* is a procedure through which researchers 1) choose a round number from the reasonable range (neighborhood) at the lower end of the raw scale to serve as the conceptual minimum, $c_n$ (Eq. 7), to anchor the lower end; 2) choose a round number from the reasonable neighborhood at the higher end to serve as the conceptual maximum, $c_x$ (Eq. 7), to anchor the higher end. As such, neighborhood rounding is a tool that helps researchers to execute conceptual anchoring.

In the example above, to anchor variable age, Zhao and Zhang (2014) selected 0 and 100 based on comparability, consistency, and convertibility, rather than 18 and 83 based on the data under examination. The researchers had other options for the upper anchor, such as the age of the oldest person known to have lived, or the average life expectancy, in the world, in the country, or in the city where the sample came from. The results based on round-number anchors that the authors chose are easier to convert and compare to other numbers based on other scales. Round numbers are more convenient for readers, reviewers, and other researchers. They are therefore preferred as the default frames of reference for comparisons over time and cross studies.

These explanations contain some main elements of what we call nine C principles of percentage scales, which will be discussed further in a later paper in this series.

## Conclusion and Anticipation

This study contents that percentage scale (*ps*) and percentage coefficient ($b_p$) may be traced to digital (by ten fingers) counting and percentage (1 for 100%) thinking of the homo sapiens ancestors millenniums back. The study then retraces and reviews three direct roots, i.e., immediate up-streams, of *ps* and $b_p$. One is the POMP theory and technique that convert close-end numerical variables to 0-100 percent scales and calculate POMP regression coefficient based on the scales (Cohen et al., 1999). The second is the min-max normalization as a data preprocessing routine for data mining, machine learning, and AI engineering (Ding et al., 2001; Han et al., 2012; Jain et al., 2005; Shalabi et al., 2006). The third is the percentage theory, techniques, and tests that began in Wisconsin in the 1980s, evolved in Pennsylvania, North Carolina and Hong Kong through the 2010s, and continue to evolve, expand, and might explode in Macau in the 2020s (Han et al., 2023; Jiang et al., 2021; Thorson et al., 1987; Thorson & Zhao, 1997/2014; Zhao, 1989, 1997; Zhao et al., 2024; Zhao & Zhang, 2014; Zhao et al., 1994).

Now we turn from upstream to downstream. Several series of manuscripts are being written or planned to explore the extensions and applications of percentage scale and percentage coefficient. A "Theory Paper" series will follow Zhao et al. (2024) as Paper 1 and this manuscript as Paper 2 to further develop the concepts and theory. A "Tool Kit" series will design and develop data analysis tools based on the *percentage theory* developed in the theory papers. A "Computational Document" series will record the computational algorithms designed to execute the theories and tools developed in the "Theory Paper" and "Tool Kit" series. An "Example Study" series will showcase the possible applications of the theory, tools, and algorithms developed in the other three series.